\documentclass[twocolumn,a4paper,superscriptaddress,floatfix,showpacs]{revtex4}
\usepackage{amsmath,amssymb,epsfig,epstopdf,color,textcase}
\usepackage{natbib}

\begin{document}

\title{Magnetostriction and ferroelectric state in AgCrS$_2$}

\author{Sergey V.~Streltsov}
\affiliation{M.N. Miheev Institute of Metal Physics of Ural Branch of Russian Academy of Sciences, 620137, Ekaterinburg, Russia}
\affiliation{Ural Federal University, Mira St. 19, 620002 Ekaterinburg, Russia}
\email{streltsov@imp.uran.ru}
\author{Alexander I. Poteryaev}
\affiliation{M.N. Miheev Institute of Metal Physics of Ural Branch of Russian Academy of Sciences, 620137, Ekaterinburg, Russia}
\affiliation{Institute of Quantum Materials Science, Bazhova St. 51, Ekaterinburg 620075, Russia}
\author{Alexey N. Rubtsov}
\affiliation{Department of Physics, Moscow State University, Moscow 119991, Russia}
\affiliation{Russian Quantum Center, Moscow 143025, Russia}


\date{\today}

\begin{abstract}
The band structure calculations in the GGA+U approximation show
the presence of additional lattice distortions in the magnetically ordered
phase of AgCrS$_2$. The magnetostriction leads to formation
of the long and short Cr-Cr bonds in the case when respective Cr ions have
the same or opposite spin projections. These changes of the
Cr lattice are accompanied by distortions of the CrS$_6$ octahedra,
which in its turn leads to development of the spontaneous electric polarization.
\end{abstract}

\maketitle

\section{Introduction}

Transition metal oxides and sulfides with delafossite structure are actively 
investigated nowadays due to diverse physical properties observed in these materials. 
For example, CuMnO$_2$ was found to show a quite strong dependence 
of magnetic properties on doping~\cite{Poienar2011,Ushakov2014}, 
CuAlO$_2$ is one of rare $p-$type transparent semiconductors~\cite{Benko1984,Laskowski2009}, 
CuFeO$_2$ was intensively studied over past years due to its multiferroicity~\cite{Ye2006}.
Another system with crystal structure closely related to
delafossites, AgCrS$_2$, was recently found to be multiferroic~\cite{Singh2009}. 
However the mechanism of the coupling between electric and magnetic characteristics
in this material is still unknown.

\begin{figure}[b!]
 \centering
 \includegraphics[clip=false,width=0.4\textwidth]{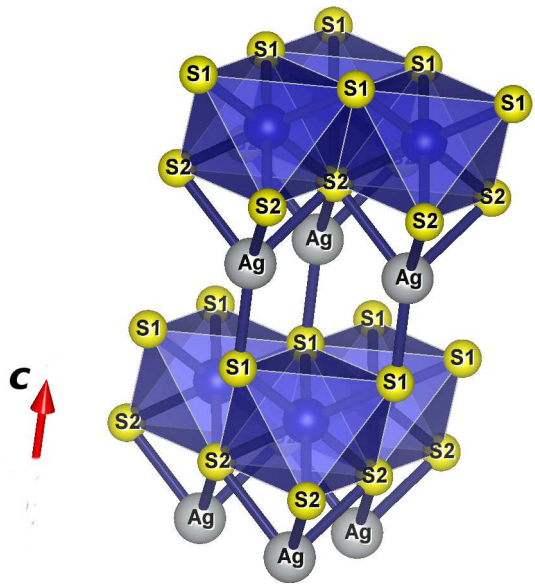}
  \caption{\label{cryst-str} (color online) Crystal structure of AgCrS$_2$. 
        Cr ions forming triangular planes are shown in blue, 
        while S and Ag ions are in yellow and grey colors, respectively. 
        It is important that there are two crystallographically different sulfur ions: 
        each Ag is connected with one S$_1$, but with three S$_2$, which results in different
        Cr-S bond lengths.}
\end{figure}

The crystal structure of AgCrS$_2$ is shown in Fig.~\ref{cryst-str}.
Triangular planes of Cr ions are stacked along $c$ direction. 
At very high temperatures it is characterized by the space group $R\bar 3m$, which 
is centro-symmetric. With decrease of the temperature to 
$T_c$=670~K it transforms to $R3m$, which is still of high symmetry 
but with lacking inversion and thus noncentro-symmetric. 
However, spontaneous electric polarization is not observed down to N\'{e}el temperature, $T_N$$\approx$41~K,
where magnetically ordered phase develops~\cite{Singh2009}. 
This low temperature phase is characterized by monoclinic noncentro-symmetric $Cm$ space group and
doubling of unit cell~\cite{Damay2011}.
Nevertheless the general triangular plane geometry is preserved and it can be regarded as distorted high symmetry.
Magnetic structure for $T<T_N$ consists of the double ferromagnetic (FM)
stripes coupled antiferromagnetically (see Fig. 4 in Ref.~\cite{Damay2011}).
This magnetic order is developed due to strong antiferromagnetic (AFM) exchange interaction
between third nearest neighbors~\cite{Ushakov2013}.

The absence of the electric polarization in $R3m$ phase is ascribed
to disorder of Ag$^+$ ions~\cite{Singh2009}, that additionally
complicates analysis of interplay between electric and magnetic
characteristics in AgCrS$_2$. Different scenarios were proposed 
to get an insight about mechanism of multiferroicity,~\cite{Damay2013,Damay2011}
but this riddle is far from being solved.

In the present paper we performed the optimization of the crystal
structure obtained previously in the experiments on powder samples
of AgCrS$_2$ and found the presence of additional lattice 
distortions in the magnetically ordered phase. 
In the direction perpendicular to the magnetic stripe (in the Cr planes) the chromium ions
with the opposite spins become closer to each other and they are moved
apart for the same spin projections.
These distortions result in the development of spontaneous electric polarization, 
which has a magnetostrictive origin therefore.

\section{Calculation details}

We used pseudo-potential Vienna {\it ab initio} simulation package (VASP) for the calculation
of electronic and magnetic properties of AgCrS$_2$~\cite{Kresse1996}.
The Perdew-Burke-Ernzerhof~\cite{Perdew1996}
version of the exchange-correlation potentials was utilized. 
The strong Coulomb correlations were taken into account via the 
GGA+U method~\cite{Anisimov1997}. The on-cite Coulomb interaction
($U$) and Hund's rule coupling ($J_H$) parameters were taken to be $U$=3.7 eV and
$J_H$=0.8 eV~\cite{Streltsov2008,Streltsov2008a}. The integration in the course of the
self-consistency was performed over a mesh of
175 {\bf k}-points in the irreducible part of the Brillouin-zone.

The electric polarization was calculated using the Berry phase 
formalism~\cite{Vanderbilt1993,Resta1994}.
The crystal structures of AgCrS$_2$ were taken from Ref.~\onlinecite{Damay2011}. 
The results presented in Sec.~\ref{two} were obtained for the data 
corresponding to $T$=10~K ($Cm$), while in Sec.~\ref{three} both $T$=10~K ($Cm$) 
and $T$=300~K ($R3m$) structures were used.

\section{\label{two}Electronic and magnetic properties and lattice distortions}

\begin{figure}[t!]
 \centering
 \includegraphics[clip=false,angle=270,width=0.5\textwidth]{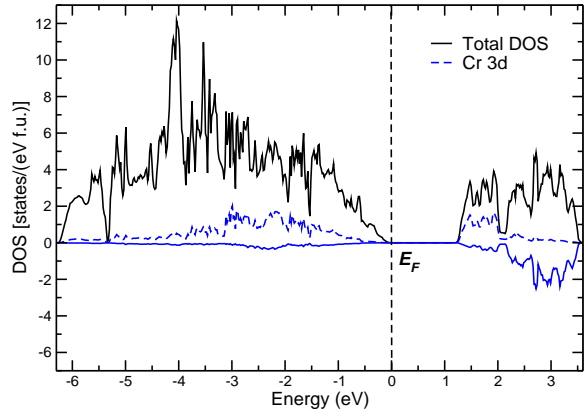}
 \caption{\label{DOS} (color online) Total and partial density of states (DOS)
         for Cr $3d$ obtained in the GGA+U calculation for the experimental double stripe
         magnetic structure. Positive and negative values of partial DOS 
         correspond to spin majority and minority. The Fermi energy is set to zero.
 }
\end{figure}

It is interesting to note that oxides based on Cr$^{3+}$ ions are usually Mott 
insulators~\cite{Matsuno2005,Sarma1996,Streltsov2008},
where top of the valence band and bottom of the conduction
band are formed by the Cr $3d$ states, and increase of the Cr oxidation state up to 4+ 
is needed to move them on the verge between Mott and charge-transfer 
regimes~\cite{Korotin1998,Streltsov2008a,Tsirlin2014,Komarek2008}. 
The density of states calculated within GGA+U approximation for experimentally observed
magnetic structure are presented in Fig.~\ref{DOS}.
An analysis of the GGA+U results shows
that Cr $3d^{\uparrow}$ states are split and occupied part lies about -4...-2~eV
while empty states that form bottom of conduction band are at 1.5...2~eV.
The minority spin states, Cr $3d^{\downarrow}$, are totally empty and located at 2...3~eV.
Magnetic moment on Cr ions is 2.9 $\mu_B$ which is in good accordance 
with the experimental value of 2.7 $\mu_B$~\cite{Damay2011}.
The top of valence band is formed predominantly by the S $3p$ states, and thus, 
AgCrS$_2$ is the charge-transfer insulator.
This is related to much larger spatial
extension of the S $3p$ orbitals compared with O $2p$. 
As a result the charge-transfer energy defined as energy costs for the
$d^np^6 \to d^{n+1}p^5$ transition is drastically decreased in sulfides~\cite{Bocquet1992}. 
The effect of the on-site Coulomb repulsion is also important since 
it splits partially occupied Cr $3d$ states and moves them away from the Fermi level. 
The band gap in AgCrS$_2$ for values of $U$ used was found to be $\sim$1.2 eV, 
which is much smaller than in oxides based on the Cr$^{3+}$ ions 
(e.g. in Cr pyroxenes the band gap is $\sim$4...5 eV~\cite{Streltsov2010}).
This can be again attributed to the decrease of charge-transfer energy in sulfides. 

In the experimental $Cm$ crystal structure each chromium atom has six in-plane
nearest neighbors: two of them lie along $b$ direction on 3.5~\AA~ distance and 
four Cr are at 3.48~\AA~ distance~\cite{Damay2011}.
The long bonded along $b$ direction Cr-Cr ions form ferromagnetically coupled chain, 
while these chains are magnetically ordered as $\uparrow \uparrow \downarrow \downarrow$
(see Fig.~4 in Ref.~\cite{Damay2011}). 
It is important to note that according to experiment the inter-chain distance remains
unaltered regardless the $\uparrow \uparrow$ or $\uparrow \downarrow$ magnetic ordering
of different chains. Naively thinking one may expect that magnetostriction differentiates
the distances between FM ($\uparrow \uparrow$) and AFM ($\uparrow \downarrow$) coupled chains and therefore the crystal structure could be somewhat different from the one 
reported previously.

In order to clarify the discrepancy between experiment and theoretical considerations
we carried out the structural optimization of the low temperature phase using GGA+U method.
The total energy calculations for experimental parameters display that 
double stripe AFM order does not correspond
to the ground state, being 1.7 meV/f.u. higher than FM state. 
The relaxation of the atomic positions and lattice parameters dramatically changes
this situation stabilizing the double stripe magnetic structure and making 
this magnetic order the lowest in total energy. 

An analysis of the relaxed atomic positions for the double 
stripe AFM structure shows that the Cr-Cr distance along
stripe stays the same, $d_{||}$=3.5~\AA, while in perpendicular
directions they change substantially: Cr-Cr bonds with the same
spin projection are stretched, $d_{\perp}^{\uparrow \uparrow}=d_{\downarrow \downarrow}$=3.52~\AA, 
and with opposite spins are shrunk, $d_{\perp}^{\uparrow \downarrow}$=3.44~\AA~
(see Fig.~\ref{P}). However the relaxation does not lead
to corrugation of the Cr planes leaving average Cr-Cr distance 
the same (3.48~\AA) in the directions perpendicular to the stripe.
Thus the effect of the magnetostriction for Cr-Cr bond lengths 
exceeds 
$\delta d_{\perp}=d_{\perp}^{\uparrow \uparrow} - d_{\perp}^{\uparrow \downarrow} \sim 0.04$~\AA~
for given $U$. 
For the FM order relaxation leads to the minor changes of the crystal structure.

It has to be noted that the effect of magnetostriction depends strongly  
on the value of the on-site Coulomb repulsion parameter $U$. 
For example, the decrease of $U$ on 1 eV results in increase of $\delta d_{\perp}$
(up to $0.05$~\AA). One of the possible explanations
could be that we gain magnetic energy moving some of the
Cr nearest neighbors closer together. For short Cr-Cr pairs
corresponding exchange interaction has to be AFM,
since the direct exchange 
should dominate over other
contributions in the edge sharing geometry~\cite{Ushakov2013}. This is 
exactly observed in the present calculations: Cr ions
having opposite spin projections are getting closer. 
Moreover, a decrease of the Coulomb interaction parameter  
increases this effect since direct exchange is inverse proportional to it, $\sim 1/U$. 
The gain in magnetic energy due to direct exchange in $\uparrow\downarrow$ pairs 
is compensated by the growth of the elastic energy and decrease
of the magnetic energy gain in $\uparrow\uparrow$  ($\downarrow \downarrow$) Cr pairs,
so that an exact value of $\delta d_{\perp}$ depends
on details on different internal parameters of the system.

It was found in Ref.~\cite{Ushakov2013} that in the LSDA (local
spin density approximation) the double stripe
AFM structure in AgCrS$_2$ is stabilized due to strong AFM 
exchange coupling between the third nearest neighbours. 
While the magnetostriction, as it was explained above, will certainly modify nearest
neighbour interaction 
the exchange to the third nearest neighbours will not be changed drastically because of
unaltered average distances in chromium plane.

\section{\label{three}Electric polarization}

For the calculation of the electric polarization ($P$) one needs
to choose a reference structure and find the difference ($\delta P$) 
in polarizations for given and reference structures~\cite{Spaldin2012}.
One usually takes the high temperature centro-symmetric lattice as the reference.
However, as it was explained above in AgCrS$_2$ the polarization appears
at the transition between two noncentro-symmetric structures, $R3m$ and $Cm$.
Therefore in spite of the fact that real reason of the absence of electric 
polarization in intermediate temperature $R3m$ structure is 
unknown~\cite{Singh2009}, we are forced to consider this structure as the 
reference.

The transition from paraelectric to ferroelectric state is accompanied by 
the transition from paramagnet to AFM with formation of the long-range magnetic order.
The paramagnetic state  having local spins and short-range magnetic correlations 
can hardly be simulated in the  GGA or GGA+U calculations. 
In principle one may model this state with such a technique as averaging over different
spin configurations~\cite{Medvedeva2002}, but even in this case this state is useless since 
for Cr$^{3+}$ with $3d^2$ electronic configuration LSDA/GGA is expected to 
give a metal and hence one cannot calculate $P$.  The nonmagnetic state could
neither be used, since the  electronic configuration of each Cr$^{3+}$ ion 
would be not $(d\uparrow)^3$,  but
 $(d\uparrow)^{1.5} (d\downarrow)^{1.5}$, which is quite unnatural
 because of the absence of the local spin. 
The isotropic AFM state (all Cr neighbours are antiferromagnetically coupled) 
which would be the best to simulate paramagnet is also impossible 
due to frustrated triangular lattice.
Therefore we simulate reference paraelectric state by the $R3m$ crystal 
structure with the FM order.


On the first stage we calculate $\delta P$ due to appearance of the
double stripe AFM order in the $R3m$ structure, relaxing
atomic positions both for the double stripe (2S) AFM and FM orders.
This results in $|\delta \vec P_{2S \to FM}|\sim 2600 \mu C/m^2$ and polarization
was found to be directed perpendicular to the Cr triangular planes (i.e.
along rhombohedral $c$ axis). We also checked that the use of the antiferromagnetically
coupled AFM ($\uparrow \downarrow \uparrow \downarrow$) chains as the reference
state does not change the result.
\begin{figure}[t!]
 \centering
 \includegraphics[clip=false,width=0.4\textwidth]{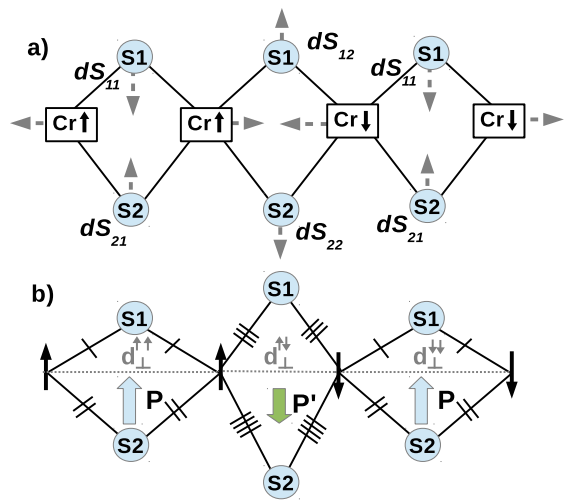}
 \caption{\label{P} (color online) 
       a) Distortions $dS_{11}$, $dS_{12}$, $dS_{21}$, $dS_{22}$ (grey dashed arrows)
          which appear in experimentally observed double stripe AFM order due to magnetostriction.  The direction perpendicular to double stipes ($\uparrow \uparrow \downarrow \downarrow$) is shown.
          Black arrows show the spin order. $dS_{11} \neq dS_{12} \neq dS_{22} \neq dS_{22}$.
       b) Distorted structure and corresponding electric polarization.
          Note that since the distortions in two neighboring CrS$_6$ octahedra are different,
          $\vec P$ and $\vec P'$ do not compensate each other.
}
\end{figure}

Since for the $R3m$ structure the lattice
parameters are known only for quite high temperature (300~K, much above
transition to ferroelectric state) and the unit cell volumes $V_{R3m}^{300K}$
and $V_{Cm}^{10K}$ are different, we repeated these calculations
for the value corresponding to the lowest in temperature structure and 
found that $|\delta \vec P_{2S \to FM}| \sim 
2800 \mu C/m^2$, i.e. practically does not change. Moreover, additional calculation 
for the $R3m$ structure,
but with fixed ionic positions (i.e. there is only electronic contribution
to the polarization) gives $|\delta \vec P^{elec}_{2S \to FM}| \sim 
1900 \mu C/m^2$. I.e. it is a redistribution of the electronic charge
density due to the double stripe AFM structure that
mainly leads to the development of the spontaneous 
electric polarization in the low temperature phase of AgCrS$_2$.
The lattice follows this tendency and changes the absolute value
of the polarization.

The double stripe magnetic structure was shown to induce ferroelectric
state due to specific lattice distortions in CdV$_2$O$_4$~\cite{Giovannetti2011},
which seems to be related with the orbital-selective
behavior~\cite{Streltsov2015MISM,Streltsov2014}.
While the crystal structure is quite different in case of AgCrS$_2$ (delafossite-like
instead of spinel), the microscopic mechanism beyond the ferroelectric 
state is similar.

In the $R3m$ crystal structure there are two inequivalent
sulfur atoms (S$_1$ and S$_2$ in Fig.~\ref{cryst-str}) with different 
chromium-sulfur bond lengths~\cite{Damay2011}. In effect 
already in the experimental $R3m$ structure 
the Cr-S$_1$-Cr and Cr-S$_2$-Cr triangles (formed by two neighboring Cr and one of
the common S) are also different in terms of angles and
bond lengths Cr-S, as shown in Fig.~\ref{P}.
The magnetostriction results in the formation
of short ($d^{\uparrow \downarrow}_{\perp}$) and long ($d^{\uparrow \uparrow}_{\perp}$)
metal-metal bonds (see Fig.~\ref{P}). 
As a result two neighboring Cr$\uparrow$-S$_1$-Cr$\uparrow$ and
Cr$\uparrow$-S$_1$-Cr$\downarrow$ triangles turn out to be differently distorted 
(the same is valid for two adjacent triangles with S$_2$ ions). 
These $dS_{11},dS_{12},dS_{21}, dS_{22}$ distortions do not compensate each other, 
which leads to non-zero electric polarization. It has to be noted, that while this
mechanism is based on specific atomic displacements, the distortions by themselves
are triggered by different charge-density distributions for the $\uparrow \uparrow$ 
($\downarrow \downarrow$) and $\uparrow \downarrow$ bonds. 

\begin{figure}[t!]
 \centering
 \includegraphics[clip=false,width=0.4\textwidth]{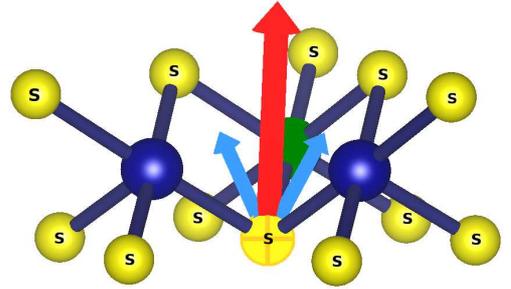}
\caption{\label{Pper} (color online) Two net electric dipole moments (shown in blue)
in two Cr$_2$S$_2$ plaquettes form net polarization (shown in red) which
is directed nearly perpendicular to the CrS$_2$ plane. The plaquettes
do not lie in the same plane (in contrast to Fig.~\ref{P}).  Blue and green balls
correspond to Cr ions with different spin projections. 
}
\end{figure}

Albeit the magnetostriction strongly modifies all bond lengths, both Cr-S$_1$-Cr and  Cr-S$_2$-Cr
triangles stay isosceles, as shown in Fig.~\ref{P}b. Therefore there 
is no component of the electric polarization along Cr-Cr bonds.
In contrast since all the displacements, $dS_{ij}$, are different 
in two neighboring plaquettes (Cr$\uparrow$-S$_1$-Cr$\uparrow$-S$_2$ and
Cr$\uparrow$-S$_1$-Cr$\downarrow$-S$_2$), 
there are two resulting polarizations $\vec P$ and $\vec P'$,
which lie in the planes of plaquettes, perpendicular to Cr-Cr
bonds. 
Each CrS$_6$ octahedron from the double chain (AFM double chain)
share its edges with two neighboring octahedra and the net
polarization is directed nearly perpendicular to the CrS$_2$ plane
as shown in Fig.~\ref{Pper}.


\section{Conclusions}

The total energy calculations in the GGA+U approximation 
for the low temperature $Cm$ crystal structure of AgCrS$_2$ compound
show an importance of atomic relaxation to set the experimentally observed magnetic order
as ground state. The relaxed atomic positions for Cr ions display the changes
of distances between FM chains that coupled ferro- and antiferromagnetically.
Therefore on the base of the first principles calculations 
we predict the existence of magnetostriction effect in this material
which was not observed in experiment~\cite{Damay2011}.
The evaluated electric polarization is mostly due to electronic rather structural origin
and it is much larger than 20 $\mu C/m^2$ measured experimentally~\cite{Singh2009}.
This discrepancy can be partially explained by the improper ferromagnetic (not paramagnetic) reference point
(which we are forced to use) to calculate initial polarization. The calculations
that treat properly the insulating nature of material at all temperatures and 
paramagnetism with local moments at high temperatures and long range magnetic order
at low temperatures can be done within DFT+DMFT method. In this case one would expect that
the calculated value of electric polarization will be reduced.
At the same time the experimental estimation of the electric polarization was obtained on the polycrystalline
samples~\cite{Singh2009}, which are known to provide substantially
smaller values than in single crystals, due to features of the pyroelectric measurements, see e.g. Refs.~\cite{Seki2008,Kimura2009,Singh2012a,Caignaert2013}.
The use of polycrystalline samples could also explain the absence of the magnetostriction effect
in experiment. Therefore, the DFT+DMFT calculations as well as refined measurements
on single crystal samples are required for further investigation of the electronic and
magnetic properties of AgCrS$_2$.

\section{Acknowledgments} 

We are grateful to A. Ushakov and especially to G. Giovannetti
for various discussions on physical properties of AgCrS$_2$ and calculation of 
the electric polarization in this material. 
A.I.P. thanks Russian Quantum Center for hospitality.
This work is supported by the Russian
Foundation for Basic Research project 13-02-00374-a and 14-02-01219-a, 
by the Ministry of Education and Science of Russia (grant MK 34432013.2) and 
Samsung corporation via GRO program.

\bibliography{../library}

\end{document}